\documentclass[journal]{IEEEtran}

\usepackage{amsmath,amsfonts}
\usepackage{algorithmic}
\usepackage{array}
\usepackage[caption=false,font=normalsize,labelfont=sf,textfont=sf]{subfig}
\usepackage{caption}
\usepackage{textcomp}
\usepackage{stfloats}
\usepackage{url}
\usepackage{verbatim}
\usepackage{graphicx}
\usepackage{tabularx}
\usepackage{multirow}
\usepackage{balance}
\usepackage{booktabs}
\usepackage{bbding}
\usepackage{contour}
\usepackage{subfig} 
\usepackage{subfloat}
\usepackage{overpic} 
\usepackage[pagebackref=false,breaklinks,colorlinks,linkcolor=black,citecolor=black]{hyperref}
\usepackage{pifont}
\newcommand{\cmark}{\ding{51}}%
\newcommand{\xmark}{\text{\ding{55}}}
\newcommand{\myPara}[1]{\vspace{5pt}\noindent\textbf{#1}}
\newcommand{\nameofmethod}{MCANet}
\newcommand{\figref}[1]{Fig.~\ref{#1}}

\newcommand{\revise}[1]{\textcolor{black}{#1}}

\begin{document}

\title{\nameofmethod{}: Medical Image Segmentation with Multi-Scale Cross-Axis Attention}

\author{Hao Shao, Quan-Sheng Zeng, Qibin Hou, \IEEEmembership{Member, IEEE}, and Jufeng Yang
\thanks{H. Shao, Q.-S. Zeng, Q. Hou, and J. Yang are with TMCC, School of Computer Science, Nankai University. }
\thanks{Q. Hou is the corresponding author (e-mail: houqb@nankai.edu.cn).}}

\markboth{Journal of \LaTeX\ Class Files,~Vol.~18, No.~9, September~2020}%
{How to Use the IEEEtran \LaTeX \ Templates}

\maketitle

\begin{abstract}
Efficiently capturing multi-scale information and building long-range dependencies among pixels are essential for medical image segmentation because of the various sizes and shapes of the lesion regions or organs.
In this paper, we present Multi-scale Cross-axis Attention (MCA) to solve the above challenging issues based on the efficient axial attention.
Instead of simply connecting axial attention along the horizontal and vertical directions sequentially, we propose to calculate dual cross attentions between two parallel axial attentions to capture global information better.
%
To process the significant variations of lesion regions or organs in individual sizes and shapes, we also use multiple convolutions of strip-shape kernels with different kernel sizes in each axial attention path to improve the efficiency of the proposed MCA in encoding spatial information.
We build the proposed MCA upon the MSCAN backbone, yielding our network, termed \nameofmethod{}.
Our \nameofmethod{} with only 4M+ parameters performs even better than most previous works with heavy backbones (e.g., Swin Transformer) on four challenging tasks, including skin lesion segmentation, nuclei segmentation, abdominal multi-organ segmentation, and polyp segmentation.
%
Code is available at \it{\url{https://github.com/haoshao-nku/medical_seg.git}}.
\end{abstract}


\section{Introduction}
\label{sec:introduction}
\IEEEPARstart{M}{edical} image segmentation is a critical and challenging research problem in medical image processing and computer vision.
This task aims to segment the parts of medical images with special significance to provide a reliable basis for clinical diagnosis and pathological research and ultimately to assist doctors in making the most accurate judgment.
It has a wide range of applications in clinical diagnosis~\cite{shi2020review,du2020medical}, computer-aided surgery~\cite{felix2021towards,huo2018synseg}, pathology analysis~\cite{chen2020unsupervised}, and other medical fields~\cite{feng2020cpfnet}.
\begin{figure}[t]
  \setlength{\tabcolsep}{0.5mm}
  \small
  \centering
  \begin{tabular}{cc}
    (a) \small Skin Lesion  & (b) Nuclei \\
    \includegraphics[width=0.24\textwidth,height=0.18\textwidth]{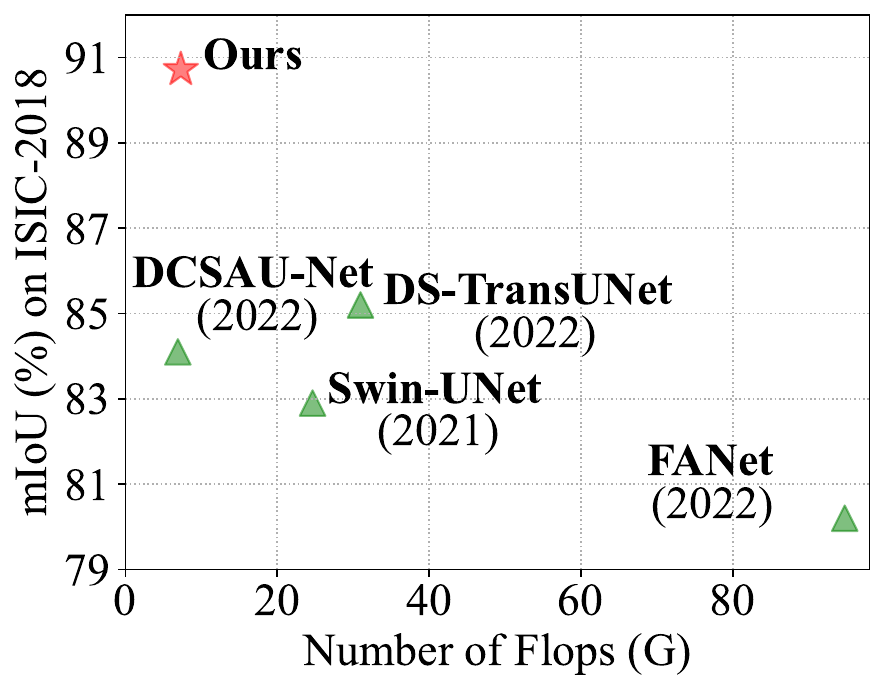}
    \put(-46.5,38.5){\scriptsize \cite{cao2021swin}}
    \put(-29,52){\scriptsize \cite{lin2022ds}}
    \put(-91,42){\scriptsize \cite{xu2023dcsau}}
    \put(-15,28.5){\scriptsize \cite{tomar2022fanet}}

    & 
    \includegraphics[width=0.24\textwidth,height=0.18\textwidth]{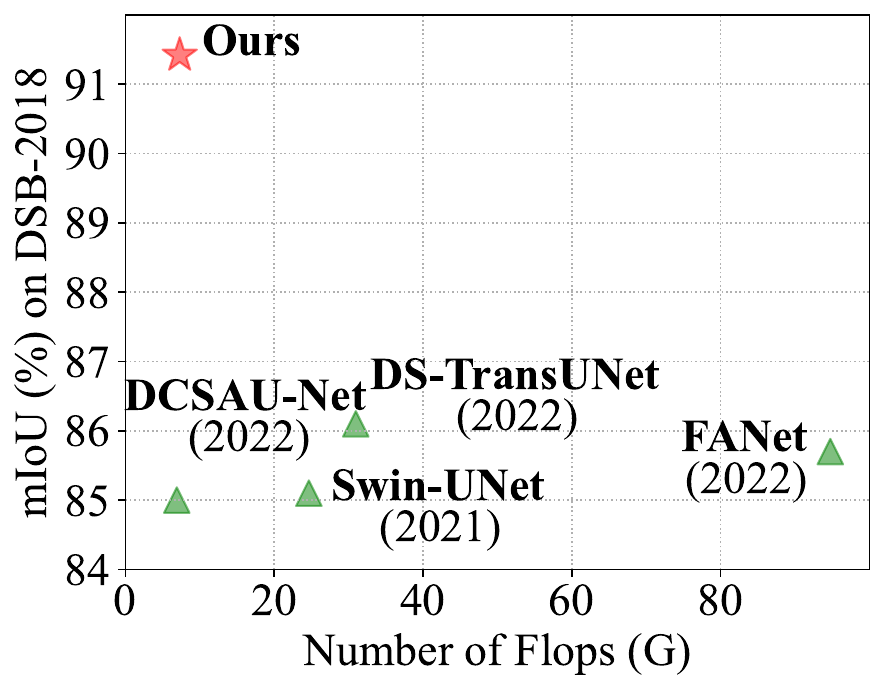}
    \put(-31,40){\scriptsize \cite{lin2022ds}}
    \put(-25,20){\scriptsize \cite{tomar2022fanet}}
    \put(-47,25.5){\scriptsize \cite{cao2021swin}}
    \put(-92,25.5){\scriptsize \cite{xu2023dcsau}}

    \\
    (c) \small Abdominal Multi-Organ & (d) Polyp    \\
    \includegraphics[width=0.24\textwidth,height=0.18\textwidth]{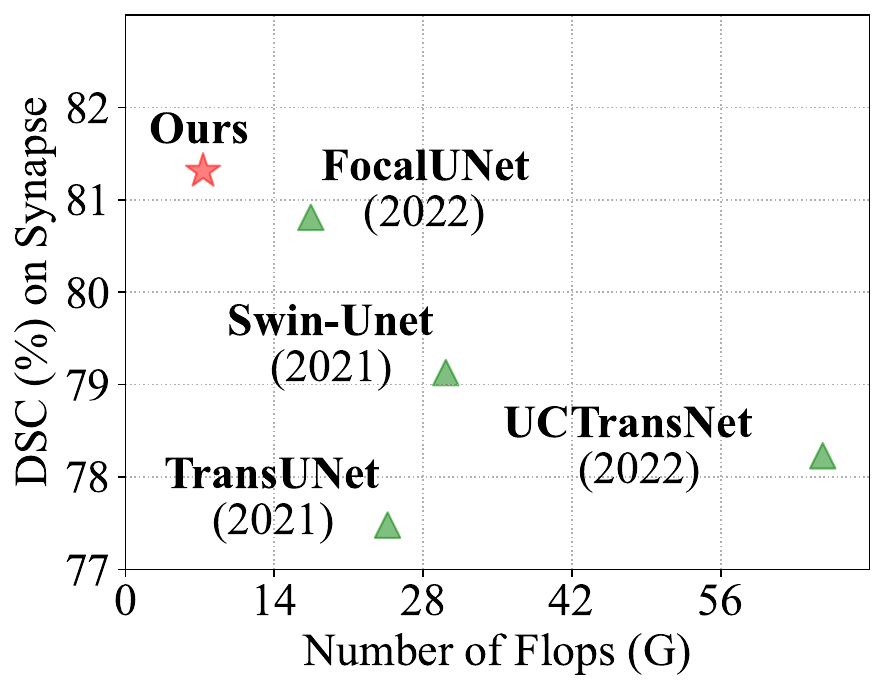}
    \put(-49,69){\scriptsize \cite{naderi2022focal}}
    \put(-62.5,48){\scriptsize \cite{cao2021swin}}
    \put(-18,34){\scriptsize \cite{wang2022uctransnet}}
    \put(-70,27){\scriptsize \cite{chen2021transunet}}

    & 
    \includegraphics[width=0.24\textwidth,height=0.18\textwidth]{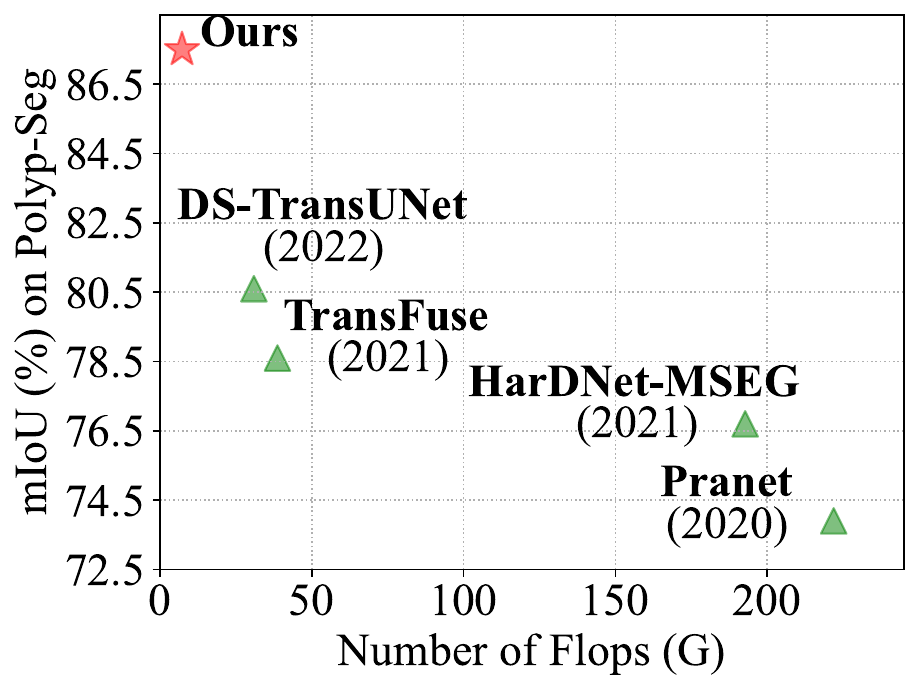}
    \put(-60,63){\scriptsize \cite{lin2022ds}}
    \put(-16,26){\scriptsize \cite{fan2020pranet}}
    \put(-57,48){\scriptsize \cite{zhang2021transfuse}}
    \put(-16,39.5){\scriptsize \cite{huang2021hardnet}}
    \\
    \end{tabular}
 \caption{Performance and Flops of our method compared to other mainstream approaches on four benchmarks. Our tiny-sized model achieves  state-of-the-art performance on all four tasks, including skin lesion segmentation, nuclei segmentation, abdominal multi-organ segmentation, and polyp segmentation, while it is more computationally efficient.}
\label{fig:flops}
\end{figure}
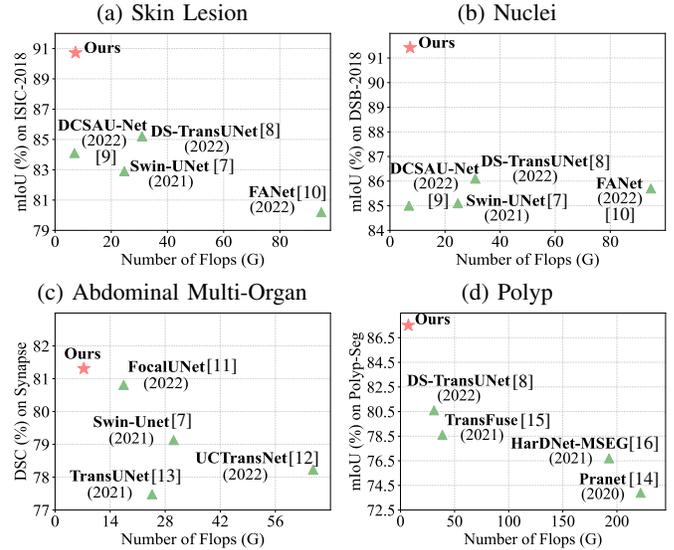

Over the past decade, with the rapid development of deep learning techniques,
neural network based approaches have gradually become the mainstream method for medical image segmentation. 
These works are primarily based on convolutional neural networks (CNNs).
In particular, U-Net~\cite{ronneberger2015u} and its variants~\cite{zhou2018unet++,xiao2018weighted,alom2018recurrent,oktay2018attention} have achieved remarkable success in recent years. 
Their success can be attributed to the encoder-decoder architecture, in which skip connections can efficiently combine the features extracted from the encoder at different scales with semantic features extracted from the decoder. 
However, the limitation of a small field of perception prevents convolutions from establishing long-distance dependencies among pixels, which has been proven essential, especially for segmentation-like tasks~\cite{hou2020strip,huang2019ccnet}. 
Due to the variability of the lesions or the organs in medical image segmentation tasks, CNN-based approaches are not well suited to handle challenges, such as variations in individual sizes, shapes, and textures~\cite{lin2022ds}. 

As an alternative to CNNs in visual recognition, Vision Transformers~\cite{dosovitskiy2020image,liu2021swin} has recently attracted significant attention. 
%
%
Since Transformers can build long-range dependencies necessary for segmentation, many works have also introduced it into the medical image segmentation task. 
For example, TransUNet~\cite{chen2021transunet} absorbs the advantages of ViT~\cite{dosovitskiy2020image} and U-Net~\cite{ronneberger2015u} to design a new network. 
Later works, such as PMTrans~\cite{zhang2021pyramid}, TransBTS~\cite{wang2021transbts}, and UNETR~\cite{hatamizadeh2022unetr}, are also proposed for medical image segmentation with different types of Transformers.
These Transformer-based methods have greatly improved the performance of previous CNNs and attain state-of-the-art results on many benchmarks.

Despite the success of Transformers in medical image segmentation, the computational complexity is a big issue when working with high-resolution images.
To alleviate this, axial attention~\cite{wang2020axial}, as an efficient way to capture global context, has been used~\cite{valanarasu2021medical,gu2020net}.
However, previous works mostly use axial attention as a tool for building long-distance relationships but focus little on the characteristics of the segmentation targets, like the sizes and the shapes of the lesion regions or the organs.
\revise{
Thus, how to efficiently encode global context and meanwhile consider the various sizes of shapes and lesion regions still deserves further exploration.
}

\begin{figure*}
  \centering
  \begin{overpic}[width=\textwidth]{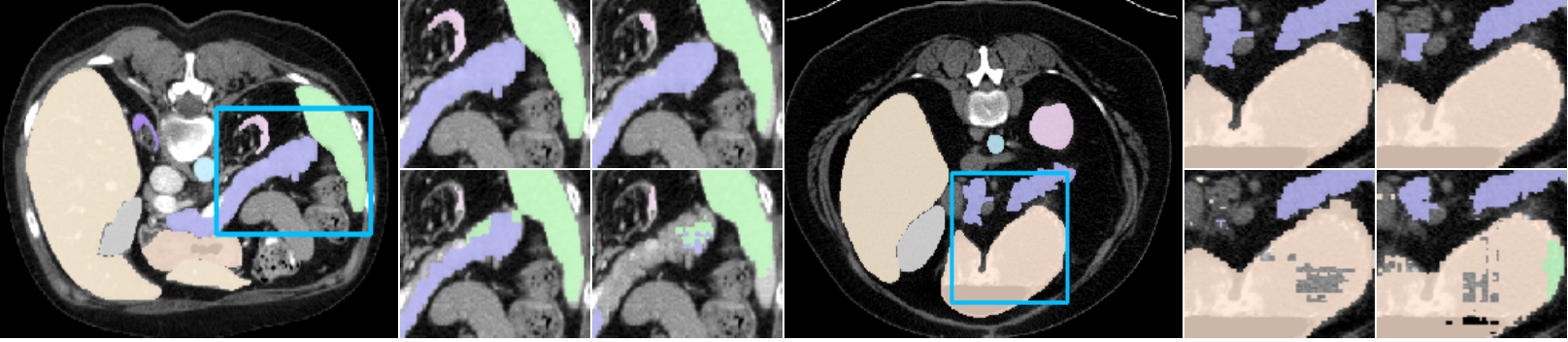}
    \put(38, 0.7){\contour{black}{\textcolor{white}{Swin-UNet}}}
    \put(38, 11.5){\contour{black}{\textcolor{white}{Ours}}}
    \put(26, 11.5){\contour{black}{\textcolor{white}{GT}}}
    \put(26, 0.7){\contour{black}{\textcolor{white}{MISSFormer}}}
    \put(88, 0.7){\contour{black}{\textcolor{white}{Swin-UNet}}}
    \put(88, 11.5){\contour{black}{\textcolor{white}{Ours}}}    
    \put(76, 11.5){\contour{black}{\textcolor{white}{GT}}}
    \put(76, 0.7){\contour{black}{\textcolor{white}{MISSFormer}}}
  \end{overpic}
  \caption{Detailed visualization of our method compared with the recently popular medical segmentation methods (e.g., MISSFormer~\cite{huang2022missformer} and Swin-UNet~\cite{cao2021swin} ) on the synapse dataset. The segmentation details produced by different methods are shown in focus in the blue rectangular box areas. Our method performs better than other methods.}
  \label{fig:detail}
\end{figure*}
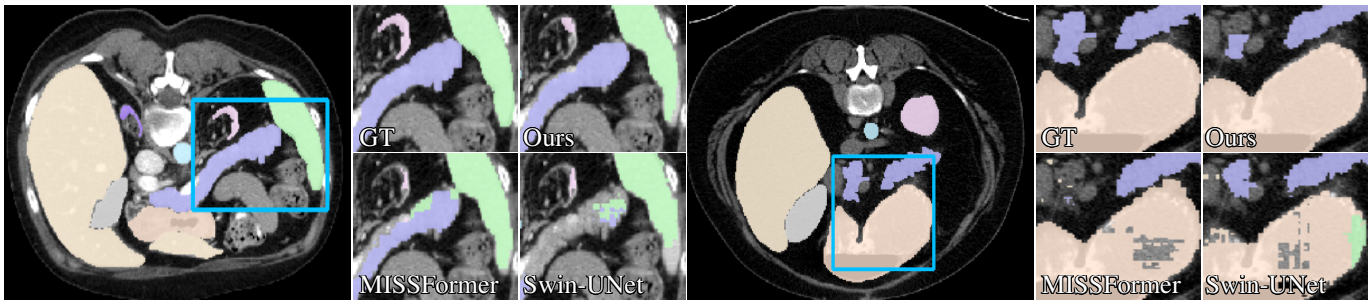

In this paper, we propose Multi-scale Cross-axis Attention (MCA). Instead of leveraging axial attention to capture global information, we renovate its design from two aspects to fit the medical image segmentation task.
First, regarding the various sizes and shapes of the
lesion regions or the organs, we attempt to involve multi-scale features in axial attention calculation using strip shape convolutions to help better locate the target regions.
%
Furthermore, rather than sequentially connecting the axial attention along the horizontal and vertical dimensions, we build dual cross attentions between the two spatial axial attentions.
This way to build interactions between the two spatial axial attentions can better use the multi-scale features and enable our method to globally identify the blurred boundaries of the target regions, which remain challenging.
%
%

%
Connecting the proposed MCA decoder to the MSCAN backbone~\cite{guo2022segnext} yields our network, dubbed \nameofmethod{}.
As shown in~\figref{fig:flops}, our proposed \nameofmethod{}, with less computational complexity, achieves the best results on a series of widely-used benchmark datasets, including skin lesion segmentation, nuclei segmentation, abdominal multi-organ segmentation, and polyp segmentation.
Some visual segmentation results are shown in \figref{fig:detail}, from which we can observe that our proposed method performs better when dealing with changes in individual size and shape, etc.
%
In summary, our main contributions can be concluded as follows:
\begin{itemize}
    \item We propose Multi-scale Cross-axis Attention (MCA), which can capture long-range dependencies and encode multi-scale information simultaneously without introducing much computational complexity.
    \item We design \nameofmethod{} based on the proposed MCA, which achieves great segmentation performance. Designing such networks is crucial to accommodate the trend of medical imaging shifting from the laboratory to the bedside.
    \item Experiments on four typical tasks show that  \nameofmethod{} outperforms previous state-of-the-art methods with fewer parameters and lower computational cost.
\end{itemize}
\section{Related Work}
This section briefly describes the network architectures and some multi-scale feature aggregation methods for medical image segmentation.
We also review some attention mechanisms that are related to our work.

\subsection{Architectures for Medical Image Segmentation}
Early networks are primarily based on convolutional neural networks (CNNs), especially U-Net~\cite{ronneberger2015u} and its variants~\cite{zhou2018unet++,xiao2018weighted,alom2018recurrent,oktay2018attention,zhou2019high} that adopt the encoder-decoder structure and have shown excellent performance.
Among them, UNet++~\cite{zhou2018unet++} uses a series of lattice-like dense skip connections to improve the segmentation accuracy.
Attention U-Net~\cite{oktay2018attention} designs a novel attention gate (AG) mechanism that enables the model to focus on targets of different shapes and sizes.
Res-UNet~\cite{xiao2018weighted} adds a weighted attention mechanism to improve the segmentation performance.
R2U-Net~\cite{alom2018recurrent} combines the advantages of residual networks~\cite{he2016deep} and U-Net~\cite{ronneberger2015u}. 
KiU-Net~\cite{valanarasu2020kiu} proposes a novel structure that exploits under-complete and super-complete features to help better segment the lesion regions with small anatomical structures.
\begin{figure*}
\centering
\includegraphics[width=\textwidth]{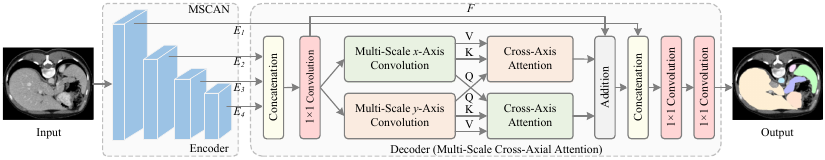}
\caption{Overall architecture of the proposed \nameofmethod{}. We take the MSCAN network proposed in SegNeXt~\cite{guo2022segnext} as our encoder because of its capability of capturing multi-scale features. The feature maps from the last three stages of the encoder are combined via upsampling and then concatenated as the input of the decoder. Our decoder is based on multi-scale cross-axis attention, which takes advantage of both multi-scale convolutional features and the axial attention.}
 \vspace{-1.0em}
\label{fig:backbone}
\end{figure*}

Another commonly used framework is based on Transformers~\cite{dosovitskiy2020image,huang2022rtnet}.
Typically, Karimi et al.~\cite{karimi2021convolution} proposed a pure Transformer-based 3D medical image segmentation model by computing the self-attention between linear embeddings of each pair of image patches.
Cao et al.~\cite{cao2021swin} proposed Swin-UNet, following Swin Transformer~\cite{liu2021swin}, which computes the self-attention within local windows to save computational complexity.
Yan et al.~\cite{yan2022after} proposed Axial Fusion Transformer UNet (AFTer-UNet), which contains a computationally efficient axial fusion layer between the encoder and the decoder to fuse inter- and intra-slice information for 3D medical image segmentation.
%
%
%


\subsection{Multi-Scale Feature Integration}

Most ViT-based medical image segmentation methods are difficult to capture multi-scale information because they mostly split the input images into fixed-size patches, thus losing useful information~\cite{shamshad2022transformers}. 
To address this problem, Zhang et al.~\cite{zhang2021pyramid} proposed a Pyramidal Medical Transformer (PMTrans), which uses multi-resolution attention to capture correlations at different image scales based on a pyramidal structure~\cite{ghiasi2016laplacian}.
Ji et al.~\cite{ji2021multi} proposed a Multi-Compound Transformer (MCTrans),
which not only learns the feature consistency of the same categories but also captures the correlations between different categories.
In order to handle volumetric data directly, Hatamizadeh et al.~\cite{hatamizadeh2022unetr} proposed a ViT-based architecture (UNETR).
UNETR includes a pure Transformer as an encoder to learn the sequence representations of the input volume, followed by a CNN-based decoder via skip to compute the final segmentation results. 
However, one of the drawbacks of UNETR is its considerable  computational complexity when dealing with sizeable 3D input volumes.
To handle this, Xie et al.~\cite{xie2021cotr} proposed a more computationally friendly self-attention module that computes multi-scale features only over a small set.
There is also a series of works in semantic segmentation aiming at designing high-level modules to capture multi-scale features~\cite{wang2020deep,liu2019auto,ngo2019deep,zhang2020exploring}.

%

\subsection{Attention Mechanisms}

Attention mechanisms play an important role in both scene image segmentation and medical image segmentation.
Based on different CNN architectures, many efforts have been made to increase the perceptual fields of networks through the self-attention mechanism~\cite{wang2018non,vaswani2017attention,wang2020dofe}.
The traditional intensive self-attention mechanism imposes heavy computational complexity and parameters and, therefore, is difficult to apply to real medical application scenarios.
Several works have proposed advanced solutions to this problem~\cite{huang2019ccnet,wang2020axial,tang2022matr}. 
For example, axial attention~\cite{wang2020axial} is computed within local windows sequentially along the horizontal and vertical axes to capture global context. 
CCNet~\cite{huang2019ccnet}  recursively employs self-attention along its vertical and horizontal paths for each pixel.

Furthermore, there is also a family of lightweight attention mechanisms~\cite{hu2018squeeze,chen2022aau}, which  significantly promote the development of segmentation methods.
Typically, SENet and its follow-ups~\cite{hu2018squeeze,misra2021rotate} adopt global pooling operators to capture global information.
Hou et al.~\cite{hou2020strip,hou2021coordinate} proposed to leverage strip pooling to emphasize the importance of positional information in building global attention.
FcaNet~\cite{qin2021fcanet} designs frequency-based channel attention networks to build attention among channels.
With the popularity of Transformers in computer vision, more attention mechanisms are proposed to reduce the computational complexity of self-attention.
Some of them propose to compute attention within local windows~\cite{liu2021swin,cao2021swin}.
Some works also attempt to reduce self-attention by developing efficient segmentation models~\cite{xie2021segformer,wang2021pyramid}.



\section{Method}


\subsection{Overall Architecture}
The overall architecture of our method is depicted in Fig.~\ref{fig:backbone}.
Following most previous works~\cite{guo2022segnext,chen2017deeplab}, we also use the classic encoder-decoder architecture.

\begin{figure*}
  \centering
  \setlength{\abovecaptionskip}{0pt}
  \includegraphics[width=0.9\textwidth]{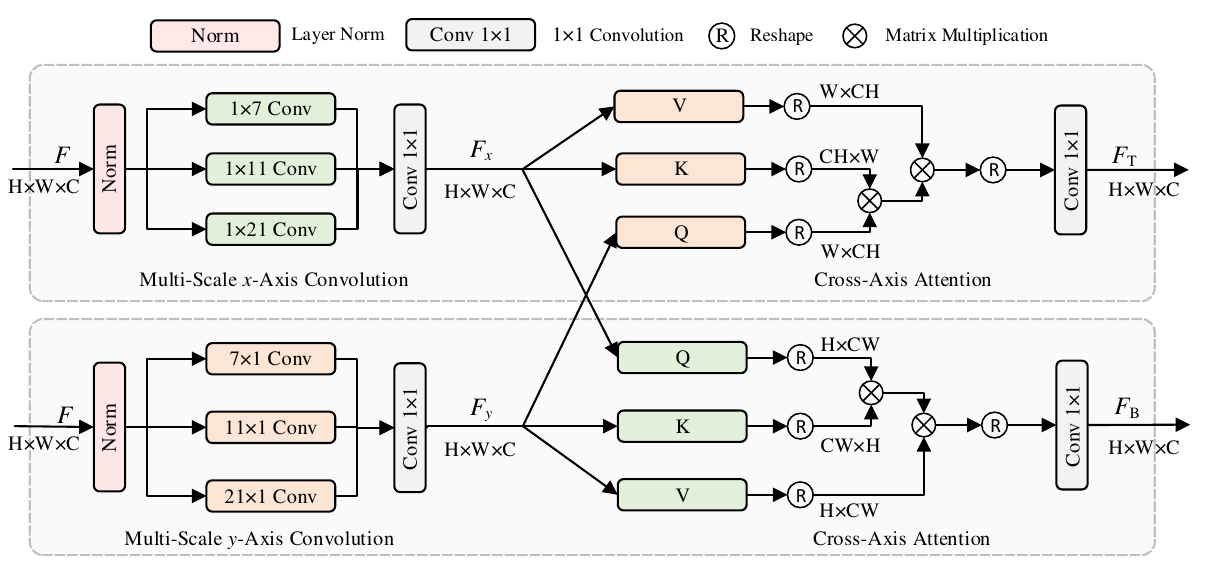}
  \caption{Detailed structure of the proposed multi-scale cross-axis attention decoder. Our decoder contains two parallel paths, each of which contains multi-scale 1D convolutions and cross-axis attention to aggregate the spatial information. \revise{Note that we do not add any activation functions in decoder.}
  }
  \label{fig:attention}
   \vspace{-1.5em}
\end{figure*}

\subsubsection{Encoder}
We adopt the recent popular computationally friendly MSCAN network from SegNeXt~\cite{guo2022segnext} as our encoder.
Unlike recent popular Transformer-based models that use self-attention to encode spatial features, SegNeXt designs a simple multi-scale convolutional attention that can efficiently capture multi-scale information.
In practice, we extract features from the four stages of the backbone,
denoted by $E_1,E_2, E_3$, and $E_4$, respectively, and feed them into the decoder.
In addition, we also attempt to use other popular encoder networks in the ablation experiment section, like
Swin Transformer~\cite{liu2021swin} and MiT from Segformer~\cite{xie2021segformer}, and report the corresponding results using them as encoders.

\subsubsection{Decoder}
Different encoder stages extract features containing different resolutions.
The features extracted from earlier stages contain local detail information, while features from deep stages possess higher-level semantic knowledge~\cite{hou2019deeply}.
Since these features have different kinds of semantic information, how to make full use of the features at different stages is crucial for the medical image segmentation task.
%
Given the feature maps $E_1, E_2, E_3$ and $E_4$ with resolutions $ \frac{\hat{H}}{4} \times \frac{\hat{W}}{4} $, $\frac{\hat{H}}{8} \times \frac{\hat{W}}{8} $, $\frac{\hat{H}}{16} \times \frac{\hat{W}}{16} $, and $ \frac{\hat{H}}{32} \times \frac{\hat{W}}{32}$, where $\hat{H}$ and $\hat{W}$ are the height and width of the input image,
our proposed multi-scale cross-axis attention decoder fuses them together via the following method.

First, feature maps $E_2, E_3$, and $E_4$ are upsampled to the same size as $E_1$ by linear upsampling, yielding $E_2^\prime,E_3^\prime$, and $E_4^\prime$.
Then, we concatenate $E_2^\prime,E_3^\prime$, and $E_4^\prime$ and feed the result into a $1\times 1$ convolution for channel dimension reduction whose output has the same channel number as $E_1$.
The output of the $1\times 1$ convolution is delivered to the multi-scale cross-axis attention,  which will be described in the next subsection.
The calculation process can be written as follows:
\begin{equation}
\mathrm{attn}=\mathrm{MCA}(\mathrm{Conv}_{1\times 1}([E_2^\prime, E_3^\prime, E_4^\prime])),
\end{equation}
where $\mathrm{MCA}(\cdot)$ denotes the multi-scale cross-axis attention, $\mathrm{Conv_{1\times 1}}(\cdot)$ denotes $1\times 1$ convolution, and
$[\cdots]$ means the concatenation operation.
%
%

%
To take advantage of the rich boundary information from $E_1$, we first concatenate the output of our multi-scale cross-axis attention and $E_1$ along the channel dimension.
The results are delivered to a couple of $1\times1$ convolutions, followed by
a bilinear interpolation operation to generate the prediction map with the exact resolution as the input medical image.
The calculation formula can be written as follows:
\begin{equation}
\mathrm{output}=\mathcal{F}_{up}{(\mathrm{Conv}_{1\times 1}(\mathrm{Conv}_{1\times 1}([\mathrm{attn},E_1])))},
\end{equation}
where $\mathcal{F}_{up}(\cdot)$ denotes the bilinear interpolation operation.
\revise{
Regarding the final result prediction part of the network, for binary classification tasks, such as skin lesion segmentation, cell segmentation, and polyp segmentation, we use $1\times 1$ convolution with only one channel to predict the final result.
For multi-class segmentation tasks, such as abdominal organ segmentation, the channel number of the last $1\times1$ convolution is determined by the number of categories.}
Furthermore, we use cross-entropy loss to optimize the whole network.

%

%


\subsection{Multi-Scale Cross-Axis Attention}

Our multi-scale cross-axis attention is a novel way to embed multi-scale features into axial attention, which aims to better segment the lesion regions  or the organs  with various individual sizes and shapes.
Before introducing the proposed multi-scale cross-axis attention, we briefly review axial attention~\cite{ho2019axial}.

\subsubsection{Revisiting Axial Attention}
Axial attention is considered as an alternative to self-attention, which decomposes self-attention into two parts, each of which is responsible for calculating self-attention along either the horizontal or the vertical dimension. 
%
%
Based on axial attention, Axial-Deeplab~\cite{wang2020axial} aggregates features along the horizontal and vertical directions sequentially, making capturing global information possible.
%
%
Thus, axial attention is more efficient than self-attention, reducing the computational complexity from $\mathcal{O}(HW \times HW)$ to $\mathcal{O}(HW \times (H+W))$.

Despite the efficiency in capturing global context, a drawback of axial attention is that it evaluates general segmentation datasets with a large amount of data and hence can learn position bias more effectively in Key, Query and Value.
In many medical image segmentation tasks, where the dataset is relatively small, it is challenging to achieve long-distance interactions~\cite{valanarasu2021medical}, which is essential to capture spatial structures or shapes in segmentation-like tasks~\cite{wang2020axial}. 
Therefore, we build dual cross attentions between the two spatial dimensions to better take advantage of directional information extracted from the axial attentions.
In addition, inspired by SegNeXt~\cite{guo2022segnext}, we also integrate multi-scale convolutional features into axial attention to conquer the challenges mentioned in Sec.~\ref{sec:introduction}.

\subsubsection{Multi-Scale Cross-Axis Attention}

Our proposed multi-scale cross-axis attention structure is shown in Fig.~\ref{fig:attention}.
It is divided into  two parallel branches, which compute the horizontal and vertical axial attentions, respectively.
Each branch consists of three 1D convolutions with different kernel sizes encoding multi-scale contextual information along one spatial dimension, followed by cross-axis attention aggregating features along the other spatial dimension.
Our goal is to harness the power of convolutions to capture multi-scale feature representations.
This makes our method quite different from previous works using axial attention.
%
In the following, we will take the top branch in Fig.~\ref{fig:attention} as an example to explain the working mechanism behind our multi-scale cross-axis attention.

Given the feature map $F$ with shape $H \times W \times C$\footnote{Here, $H = \frac{\hat{H}}{4}$ and $W = \frac{\hat{W}}{4}$.}, i.e., the combination of the feature maps from the last three stages of the encoder, we use three parallel 1D convolutions to encode $F$, the outputs of which are fused via summation and sent into a $1\times1$ convolution.
The formulation can be written as follows:
\begin{equation}
F_{x} =  \mathrm{Conv_{1\times 1}}(\sum_{i=0}^{2}\mathrm{Conv1D}_{i}^x(\mathrm{Norm}(F))),
\end{equation}
where $\mathrm{Conv1D}_{i}^x(\cdot)$ denotes a 1D convolution along the $x$-axis dimension, $\mathrm{Norm}(\cdot)$ is a layer normalization, and $F_{x}$ is the output.
For the kernel sizes of the 1D convolutions, we, following SegNeXt~\cite{guo2022segnext}, set them to $1\times7$, $1\times11$, and $1\times21$, respectively.
For the bottom branch, the corresponding output $F_{y}$ can be attained by:
\begin{equation}
F_{y}=\mathrm{Conv_{1\times 1}}(\sum_{i=0}^{2}\mathrm{Conv1D}_{i}^y(\mathrm{Norm}(F))).
\end{equation}

Given $F_{x}$ in the top branch, we deliver it into a $y$-axis attention.
To better use the multi-scale convolutional features from both spatial directions, we propose to compute the cross attention between $F_{x}$ and $F_{y}$.
Specifically, we take $F_{x}$ as the key and value matrices while view $F_{y}$
as the query matrix.
%
The calculation process can be written as follows:
\begin{equation}
F_T = \mathrm{MHCA}_y(F_{y},F_{x},F_{x}),
\end{equation}
%
where $\text{MHCA}_y(\cdot, \cdot, \cdot)$ denotes multi-head cross-attention along the $x$ axis.
For the bottom branch, we use a similar way to encode the context along the $y$-axis direction as follows:
\begin{equation}
F_B = \mathrm{MHCA}_x(F_{x},F_{y},F_{y}),
\end{equation}
where $\text{MHCA}_x(\cdot, \cdot, \cdot)$ denotes multi-head cross-attention along the $y$ axis.


%
%
Given $F_{T}$ and $F_{B}$, the output of the proposed multi-scale cross-axis attention can be formulated as:
\begin{equation}
F_{\text{out}} =\mathrm{Conv_{1\times 1}}(F_{T}) + \mathrm{Conv_{1\times 1}}(F_{B}) + F.
\end{equation}
In Table~\ref{tab:details}, we show more network details on the hyper-parameters we use for three variants of our \nameofmethod{}.

\subsubsection{Advantages} 
Compared to other methods, our proposed multi-scale cross-axis attention  has the following main advantages:
\begin{itemize}
    \item We introduce lightweight multi-scale convolutions, which is an effective way to handle the various sizes and shapes of the lesion regions or the organs;
    \item Unlike most previous works, we do not directly apply the axial attention in our method to capture global context. Instead, we propose to build interactions between two parallel axial attentions, and a more efficient way to take advantage of both the multi-scale features and the global context;
    \item Our decoder is lightweight. As shown in Tab.~\ref{tab:details}, the number of parameters is only 0.14M for our tiny-sized model, which is more suitable for practical application scenarios.
\end{itemize}



\begin{table}[t!]
  \centering
  \small
  \setlength\tabcolsep{3.5pt}
  \caption{
  Network details for three variants of~\nameofmethod{}.
  \revise{`H' is the number of heads.
  `C' represents the number of channels after using $1\times 1$ convolution for channel dimension reduction in decoder.
  }}
  \begin{tabular}{ccccccccc} \toprule
     & \multicolumn{2}{c}{Encoder} & \multicolumn{3}{c}{Decoder} \\ \cmidrule(lr){2-3}\cmidrule(lr){4-6}
    Variants & Channels &\revise{Blocks} & C & H & Params (M)  \\ \midrule

    \nameofmethod{}-T &(32,64,160,256)   &\revise{(3,3,5,2)}    & 64 &8  & 0.14   \\ 
    \nameofmethod{}-S &(64,128,320,512)  &\revise{(2,2,4,2)}    &128 &8  & 0.55    \\ 
    \nameofmethod{}-B &(64,128,320,512)  &\revise{(3,3,12,3)}    &128 &8  & 0.55    \\ \bottomrule
\end{tabular}
 \vspace{-1em}
  \label{tab:details}
\end{table}

\section{Experiments}
This section evaluates our proposed method on four challenging tasks: skin lesion segmentation, nuclei segmentation, abdominal multi-organ segmentation, and polyp segmentation on widely used datasets. 
We also provide ablation analysis to help readers understand how each component in our method contributes to the segmentation performance.

\subsection{Datasets}

\myPara{Skin Lesion  Segmentation:} 
We select the dataset from the ISIC-2018 challenge for skin lesion segmentation.
ISIC-2018 is one of the most representative and challenging two-dimensional skin lesion segmentation datasets in computer-aided diagnosis, with 2594 images and fine annotations. 
They are randomly divided into a training set and a test set, where the training set consists of 2074 images and the test set consists of 520 images.
The challenges behind this dataset can be summarized as follows: low contrast between the lesion area and the surrounding skin, blurred lesion boundaries, and high variation between the melanoma classes.

\myPara{Nuclei Segmentation:} For the task of nuclei segmentation, we report the results on the DSB2018 dataset from the 2018 Data Science Bowl challenge. 
This dataset contains a total of 670 images with accurate annotations.
Of these, 80$\%$ are used for training, 10$\%$ for validation and 10$\%$ for testing.
The cell segmentation task has two main challenges.
First, pathology images need professional pathologists to accurately annotate, and the dataset annotation cost is high.
So, the amount of data is relatively small and overfitting often occurs when using large-scale complex networks.
Second, there is a boundary blur between cells.


\myPara{Abdominal Multi-Organ Segmentation:} For abdominal multi-organ segmentation task, we report results on the Synapse dataset proposed by TransUNet~\cite{chen2021transunet}. 
This dataset uses 30-segment abdominal CT scans from the MICCAI 2015 challenge and contains 3779 axial contrast-enhanced clinical CT images of the abdomen. 
The dataset contains eight classes of abdominal organs: aorta, gallbladder, left kidney, right kidney, liver, pancreas, spleen, and stomach.
According to the TransUNet~\cite{chen2021transunet} setup, we randomly divide the 30-segment CT scans into training and test sets based on a ratio of 3:2.
The challenges for the abdominal multi-organ segmentation task are mainly as follows: 
Significant individual differences between organs in the abdominal organ region, like the sizes and shapes and blurred boundaries between organs.


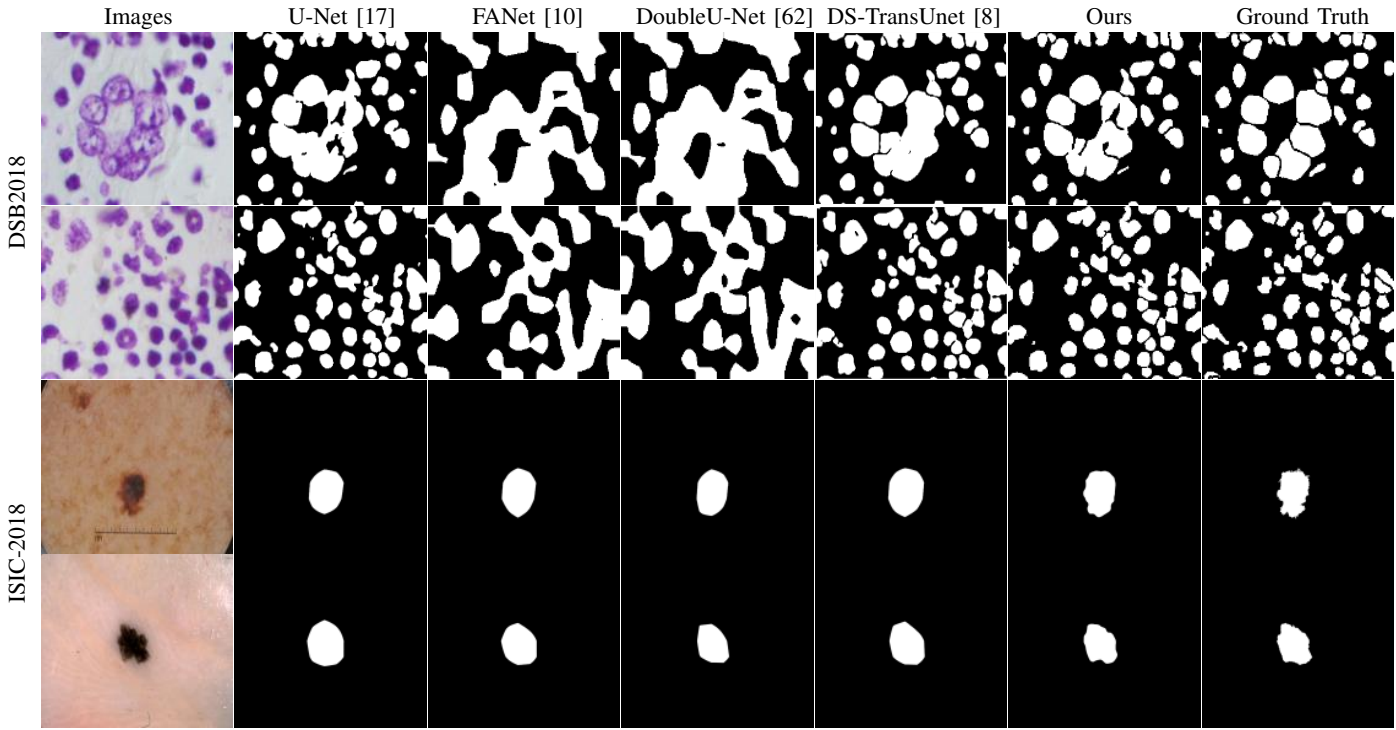
\begin{figure*}
  \centering
  \small
  \setlength{\abovecaptionskip}{8pt}
  \begin{overpic}[width=\textwidth]{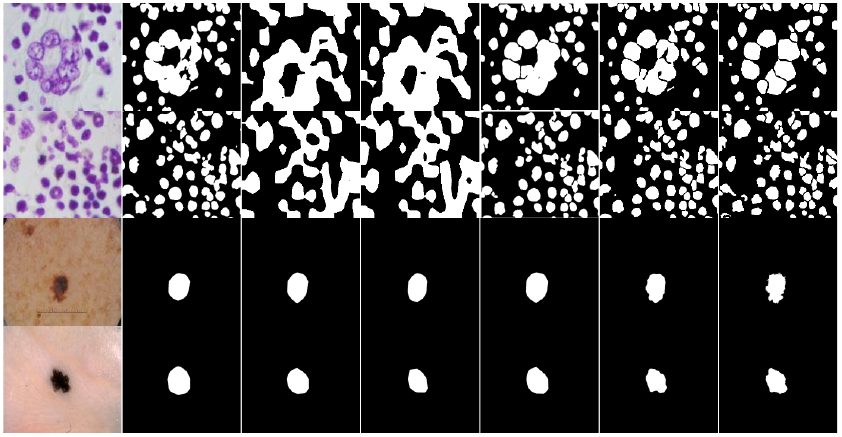}
    \put(5, 52){{\textcolor{black}{Images}}}
    \put(18.5, 52){{\textcolor{black}{U-Net~\cite{ronneberger2015u}}}}
    \put(32, 52){{\textcolor{black}{FANet~\cite{tomar2022fanet}}}}
    \put(44, 52){{\textcolor{black}{DoubleU-Net~\cite{jha2020doubleu}}}}
    \put(58, 52){{\textcolor{black}{DS-TransUnet~\cite{lin2022ds}}}}
    \put(77, 52){{\textcolor{black}{Ours}}}
    \put(88, 52){{\textcolor{black}{Ground Truth}}}

    \put(-2,35.5){\rotatebox{90}{DSB2018}}
    \put(-2,9.5){\rotatebox{90}{ISIC-2018}}
    
  \end{overpic}
  \caption{Segmentation results of different methods on the DSB2018 dataset and ISIC-2018 dataset.}
   \vspace{-1em}
  \label{fig:dsbisic}
\end{figure*}

\myPara{Polyp Segmentation:} 
For the task of polyp, we report results on the same benchmarks as used in Pranet~\cite{fan2020pranet}, which contain five commonly used datasets, including Kvasir-SEG, CVC-ClincDB, CVC-ColonDB, EndoScene, and ETIS.
Among them, 900 images from Kvasir-SEG and 550 images from ClinicDB constitute the training set. 
The test set include 100 images from Kvasir-SEG, 62 images from CVC-ClincDB, 380 images from CVC-ColonDB, 60 images from EndoScene, and 196 images from ETIS.
%
%
Difficulties for this task are listed below: polyps vary dramatically in size and shape; the boundaries around the polyps are often blurred.

\begin{table}[!tp]
  \centering
  \small
  \renewcommand\arraystretch{1}
  \setlength\tabcolsep{2pt}
  \caption{Comparison with the current state-of-the-art methods on the ISIC-2018 dataset and DSB2018 dataset.}
  \begin{tabular}{lccccccccccccccc}
  \toprule
  {} & Params & Flops &\multicolumn{2}{c}{ISIC-2018} &\multicolumn{2}{c}{DSB2018} \\ 
  \cmidrule(lr){4-5}\cmidrule(lr){6-7}
  Method & (M) & (G) & mIoU & F1-score     & mIoU   & F1-score   \\ \midrule
    U-Net~\cite{ronneberger2015u}   &24.56 &38.26  &80.20  &87.40 &80.80   &88.70    \\
    FANet~\cite{tomar2022fanet}    &7.72  &94.75  &80.23    &87.31   &85.69   &91.76\\ 
    DoubleUNet~\cite{jha2020doubleu} &29.30  &107.90  &82.12   &89.62  &84.07   &91.33 \\
    DCSAU-Net~\cite{xu2023dcsau}  &2.60  &6.91 &84.10  &-  &85.00  &-\\ 
    DS-TransUnet~\cite{lin2022ds}   &287.75 &51.09 &85.23   &91.32  &86.12   &92.19  \\ 
    \midrule
    Swin-T+MCA  &27.67 &26.93 &88.08 &91.27 &90.95 &91.58   \\
    \nameofmethod{}-T        &4.04   &7.30   &90.40   &92.93 &91.16   &91.76    \\ 
    \nameofmethod{}-S        &13.98  &23.41  &90.69   &93.04 &91.43   &92.02    \\ 
    \nameofmethod{}-B        &26.80  &36.68  &90.88   &93.18
    &91.48   &92.07\\ \bottomrule
\end{tabular}
 \vspace{-1.5em}
\label{isic-2018}
\end{table}

\subsection{Implementation Details}
We use Pytorch to implement our \nameofmethod{}  for all experiments.
\revise{For skin lesion segmentation, nuclear segmentation, and polyp segmentation experiments, our implementation is based on mmsegmentation~\cite{contributors2020mmsegmentation}.
The input image size is $512\times 512$.
We train our models for 80k iterations using the AdamW optimizer with an initial learning rate of 0.0002, momentum of 0.9, and weight decay of 1e-4.
Our implementation for the abdominal multi-organ segmentation is based on mmsegmentation and TransUNet~\cite{chen2021transunet}.
For a fair comparison, we use the same experimental setup as TransUNet. 
The input image size is $224\times 224$, and we use the SGD optimizer and train the models for 14K iterations.
The learning rate is set to 0.01, the momentum to 0.9, and the weight attenuation to 1e-4.
The batch size is set to 16 for all experiments.}
We use a NVIDIA RTX 3090 GPU for training.
Three versions of \nameofmethod{} are provided: \nameofmethod{}-T, \nameofmethod{}-S, and \nameofmethod{}-B.

For the nuclei and skin lesion segmentation tasks, we follow the settings of FANet~\cite{tomar2022fanet} using mIoU and F1-score as evaluation metrics.
For the abdominal multi-organ segmentation task, we follow the settings of Swin-UNet~\cite{cao2021swin} using DSC and HD as evaluation metrics.
For the polyp segmentation task, we use mIoU and mDice as evaluation metrics according to DS-TransUnet's~\cite{lin2022ds}.
We use mmsegmentation's self-contained calculation method to compute the number of parameters and flops based on an input image size of $512\times 512$.
%
%

\subsection{Comparison with state-of-the-art Methods}

\myPara{Results on ISIC-2018:} 
For ISIC-2018 dataset, our comparison with the current state-of-the-art methods is shown in Table~\ref{isic-2018}.
We can observe that even our proposed \nameofmethod{}-T with around 4M+ parameters outperforms the methods listed in the table.
The visualization results are shown in the bottom part of Fig.~\ref{fig:dsbisic}.
Due to the low contrast between the lesion regions and the surrounding skin, the lesion boundaries are often blurred.
In addition, the skin lesion regions vary in size and shape.
Our method uses cross-axis attention and multi-scale convolutions and hence can handle the above challenges.
It can be seen from the final segmentation results that our proposed method can recognize the segmentation boundaries more accurately, and the shapes are also close to the annotations.

%

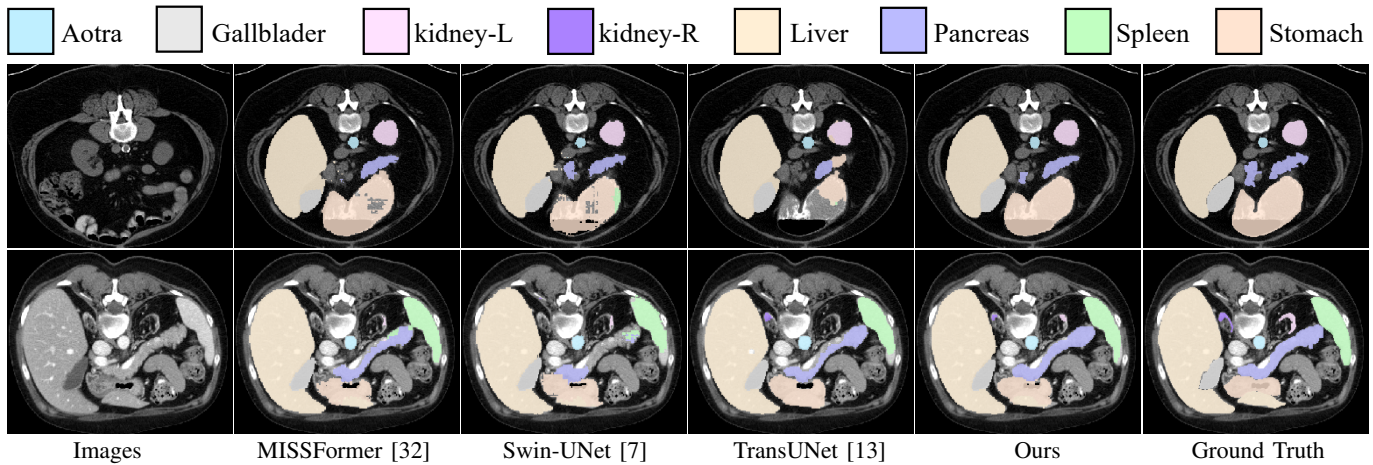
\begin{figure*}[tp]
  \small
  \centering
  \begin{overpic}[width=\textwidth]{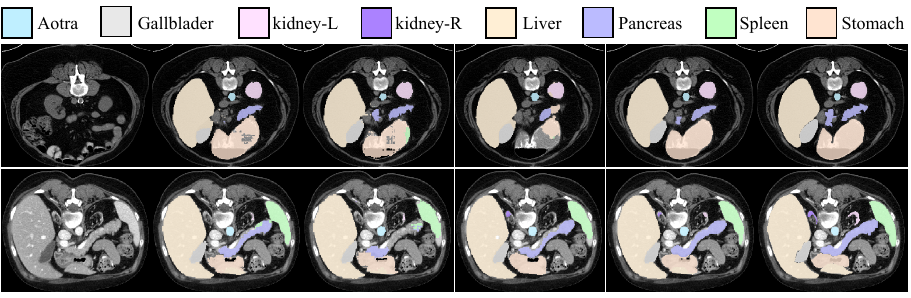}
    \put(5, -1.8){{\textcolor{black}{Images}}}
    \put(18.5, -1.8){{\textcolor{black}{MISSFormer~\cite{huang2022missformer}}}}
    \put(36.5, -1.8){{\textcolor{black}{Swin-UNet~\cite{cao2021swin}}}}
    \put(53.4, -1.8){{\textcolor{black}{TransUNet~\cite{chen2021transunet}}}}
    \put(74, -1.8){{\textcolor{black}{Ours}}}
    \put(87, -1.8){{\textcolor{black}{Ground Truth}}}

  \end{overpic}
  \vspace{-1pt}
  \caption{The segmentation results of different methods on the Synapse dataset.}
   \vspace{-1em}
  \label{fig:synapse_visual}
\end{figure*}

\begin{table*}[t!]
    \centering
    \caption{Comparison with the current state-of-the-art methods on the Synapse dataset.
    \revise{The input image size of the methods listed in the following table is ${224\times 224}$ during the experiment, except for Attention U-Net, whose input image size is ${160\times 160}$.
    }
    }
    \small
    \setlength\tabcolsep{2pt}
    \begin{tabular}{lcccccccccccc} \toprule
    Methods & Params (M) & Flops (G) &  DSC ($\uparrow$) & HD ($\downarrow$) & Aorta& Gallbladder& Kidney-L & Kidney-R & Liver& Pancreas& Spleen& Stomach\\ 
    \midrule
    U-Net~\cite{ronneberger2015u}  &24.56 &38.26  &74.68  &36.87  &87.74  &63.66  &80.60  &78.19  &93.74  &56.90 &85.87  &74.16        \\
    AttnUNet~\cite{schlemper2019attention} &34.90  &101.90  &75.57  &36.97  &55.92  &63.91  &79.20  &72.71  &93.56  &49.37 &87.19  &74.95        \\
    TransUNet~\cite{chen2021transunet} &105.28 &24.66 &77.48  &31.69  &87.23  &63.13  &81.87  &77.02  &94.08  &55.86 &85.08  &75.62        \\  
    Attention U-Net~\cite{oktay2018attention} &25.09 &40.07  &77.77 &36.02 &89.55 &68.88 &77.98 &71.11 &93.57 &58.04 &87.30 &75.75 \\
    Swin-UNet~\cite{cao2021swin}  &149.22 &30.14   &79.13  &21.55  &85.47  &66.53  &83.28  &79.61  &94.29  &56.58 &90.66  &76.60        \\
    UCTransNet~\cite{wang2022uctransnet} &65.6 &63.2   &78.23 &26.75  &-  &-  &-  &-  &-  &- &-  &-        \\
    FocalUNet~\cite{naderi2022focal} &44.25 &17.44  &80.81  &20.66  &85.74  &71.37  &85.23  &82.99  &94.38  &59.34  &88.49  &78.94 \\
    MISSFormer~\cite{huang2022missformer}  &42.46 &54.46  &81.96 &18.20 &86.99 &68.65 &85.21 &82.00 &94.41 &65.67 &91.92 &80.81 \\
    \midrule
    Swin-T + MCA  &27.67 &26.93  &80.29 &22.15 &86.71 &70.03 &84.09 &80.47 &94.11 &60.39 &89.49 &77.00   \\
    \nameofmethod{}-T &4.04    &7.30  &81.31  &24.59   &86.62  &70.66  &82.65  &80.47  &94.91  &66.24 &87.89  &81.02 \\
    \nameofmethod{}-S   &13.98 &23.41 &81.64 &18.75 &87.67 &67.66 &83.35 &80.60 &95.36 &64.21 &90.89 &83.38\\
    \nameofmethod{}-B &26.80   &36.68  &83.52  &15.82   &88.35  &71.47  &87.30  &84.53  &95.19  &65.18 &91.75  &84.42 \\
    
    \bottomrule
    \end{tabular}
     \vspace{-1em}
    \label{synapse}
\end{table*}




\myPara{Results on DSB2018:}
For DSB2018 dataset, we compare our method with a series of popular approaches and show the results in Table~\ref{isic-2018}.
Our work provides three versions of~\nameofmethod{}, all of which are superior to the methods listed in the table. 
We also visualize segmentation results in the top part of Fig.~\ref{fig:dsbisic}.
%
The segmentation results show that our proposed \nameofmethod{} can predict the nucleus boundaries more accurately than other methods.

\myPara{Results on Synapse:} 
We compare \nameofmethod{} with the state-of-the-art approaches on the Synapse dataset and report the results in Table~\ref{synapse}.
It is worth mentioning that the HD scores by our proposed \nameofmethod{}-B is only 15.82, which is much smaller than that of other methods.
Fig.~\ref{fig:synapse_visual} shows segmentation results.
It can be seen from figure that our proposed method can identify the boundaries of different organs more accurately.
This is mainly attributed to the combination of cross-axis attention and multi-scale convolutions.

\myPara{Results on Polyp Segmentation task:}  
We also report results on polyp segmentation task as shown in Table~\ref{POLYP}.
%
%
We can observe that~\nameofmethod{} 
outperforms the methods listed in the table. 
On average, our \nameofmethod{}-T model with around 4M parameters even performs better than DS-Transunet-L, which has more than 200M parameters, 50 times larger than \nameofmethod{}-T.
%

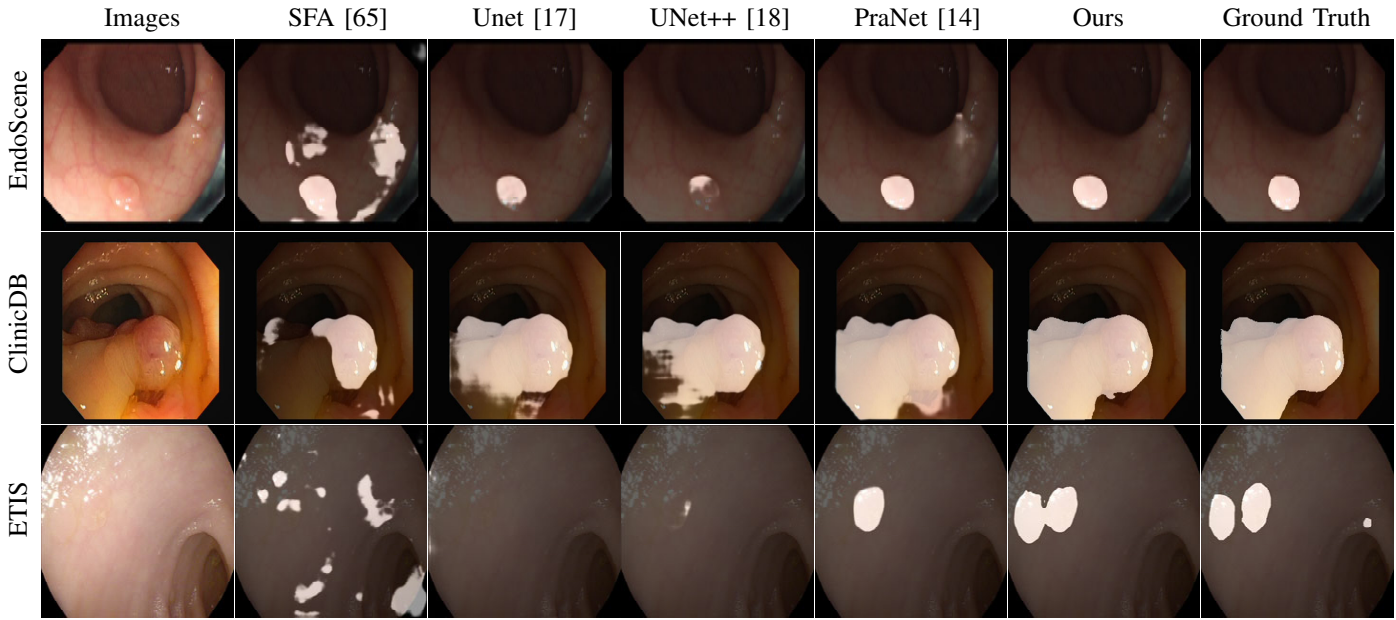
\begin{figure*}
  \centering
  \begin{overpic}[width=\textwidth]{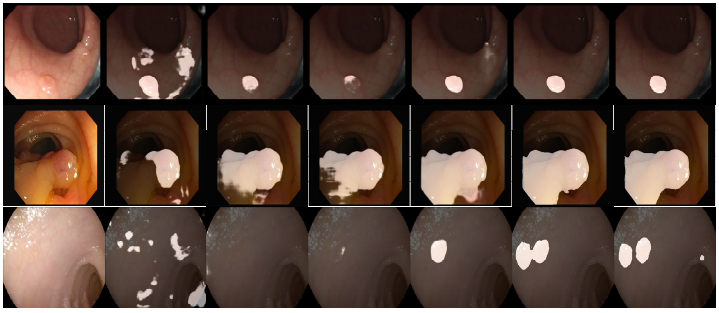}
    \put(5, 44){{\textcolor{black}{Images}}}
    \put(18.5, 44){{\textcolor{black}{SFA~\cite{fang2019selective}}}}
    \put(32, 44){{\textcolor{black}{Unet~\cite{ronneberger2015u}}}}
    \put(45, 44){{\textcolor{black}{UNet++~\cite{zhou2018unet++}}}}
    \put(60, 44){{\textcolor{black}{PraNet~\cite{fan2020pranet}}}}
    \put(76, 44){{\textcolor{black}{Ours}}}
    \put(87, 44){{\textcolor{black}{Ground Truth}}}

    \put(-2,32){\rotatebox{90}{EndoScene}}
    \put(-2,18.5){\rotatebox{90}{ClinicDB}}
    \put(-2,6.5){\rotatebox{90}{ETIS}}
  \end{overpic}
  \caption{The segmentation results of different methods on the polyp segmentation task.}
   \vspace{-0.8em}
  \label{fig:polyp-outputs}
\end{figure*}

%
The segmentation results of different methods are shown in Fig.~\ref{fig:polyp-outputs}.
The color of the polyps in the endoscopic images is very similar to that of the background, making the borders challenging to detect.
Furthermore, the sizes and shapes of the polyps are very variable.
Despite this, \nameofmethod{} equipped with cross-axis attention and multi-scale convolutions, can address the above challenges well.
%
%

\subsection{Ablation Study}
Here, we perform ablation experiments on the ISIC-2018 dataset of the skin lesion segmentation task to evaluate the importance of each component for~\nameofmethod{}.

\begin{table*}[tp]
  \centering
  \small
  \renewcommand\arraystretch{1}
  \setlength\tabcolsep{2.8pt}
  \caption{Comparison with the current state-of-the-art methods on the polyp segmentation tasks.}
  \begin{tabular}{lccccccccccccccc}
  \toprule
  {} & {} & {} &\multicolumn{2}{c}{Kvasir} &\multicolumn{2}{c}{ClinicDB} &\multicolumn{2}{c}{ColonDB} &\multicolumn{2}{c}{EndoScene} &\multicolumn{2}{c}{ETIS} &\multicolumn{2}{c}{Average} \\ \cmidrule(lr){4-5}\cmidrule(lr){6-7}\cmidrule(lr){8-9}\cmidrule(lr){10-11}\cmidrule(lr){12-13}\cmidrule(lr){14-15}
  Methods & Params (M) & Flops (G) & mDice   & mIoU    & mDice   & mIoU     & mDice   & mIoU    & mDice   & mIoU     & mDice   & mIoU    & mDice   & mIoU \\ \midrule
U-Net~\cite{ronneberger2015u}&24.56 &38.26 &81.80 &74.60 &82.30  & 75.50 &51.20 &44.40 &39.80 &33.50 &71.00 &62.60 &65.20 &58.10    \\
UNet++~\cite{zhou2018unet++}&25.09 &84.30 &82.10 &74.30 &79.40 &72.90  &48.30 &41.00 &40.10 &34.40 &70.70 &62.40 &64.10 &57.00        \\
Attention U-Net~\cite{oktay2018attention}&25.09 &40.07 &81.40 &73.00 &85.00 &78.90 &56.10 &48.40 &77.30 &68.20 &37.10 &30.50 &67.40 &59.80 \\
PraNet~\cite{fan2020pranet}&32.50 &221.90 &89.80 &84.00 &89.90 &84.90 &70.90 &64.00 &87.10 &79.70 &62.80 &56.70 &80.00 &73.90    \\
Swin-Unet~\cite{cao2021swin}&149.22 &30.14 &89.60 &83.50 &89.90 &83.60 &75.90 &66.60 &85.00 &76.40 &68.10 &58.60 &81.70 &73.70    \\
HarDNet-MSEG~\cite{huang2021hardnet}&33.80 &192.74 &91.20 &85.70 &93.20  &88.20 &73.10 &66.00 &88.70 &82.10 &67.70 &61.30 &82.80 &76.70   \\
TransFuse~\cite{zhang2021transfuse}&115.59  &38.73 & 91.80   & 86.80  & 93.40  & 88.60  & 74.40  & 67.60  & 90.40  & 83.80 & 73.70  & 66.10  & 84.70 & 78.60  \\
DS-TransUNet-B~\cite{lin2022ds}&177.44 &30.97  & 93.40  & 88.80  & 93.80  & 89.10  & 79.80     & 71.70   & 88.20 & 81.00 & 77.20 & 69.80  & 86.50  & 80.10           \\
DS-TransUNet-L~\cite{lin2022ds}&287.75 &51.09  & 93.50  & 88.90  & 93.60 & 88.70  & 79.80     & 72.20   & 91.10  & 84.60 & 76.10  & 68.70  & 86.80  & 80.60         \\ 
\midrule
Swin-T + MCA (Ours)  &27.67 &26.93 &91.59 &86.97 &97.18 &94.61 &86.77 &78.42 &93.50 &88.40 &90.58 &83.89 &91.92 &86.46 \\
\nameofmethod{}-T &4.04 &7.30 &91.49 &84.88 &97.39 &94.99 &84.94 &76.09 &95.43 &91.57 &91.79 &85.70 &92.21 &86.65\\ 
\nameofmethod{}-S &13.98 &23.41 &92.97 &87.25 &97.35 &94.93 &86.19 &77.69 &94.49 &90.00 &92.01 &86.04 &92.60 &87.18 \\
\nameofmethod{}-B &26.80 &36.68 &94.85 &90.42 &97.04 &94.37 &85.79 &77.16 & 95.07 &90.97 &92.16 &86.26 &92.98 &87.84   \\
\bottomrule 
\end{tabular}
 \vspace{-1em}
\label{POLYP}
\end{table*}

\begin{table}[!tp]
    \centering
    \small
    \centering
    \caption{Ablation study of the  Multi-scale Cross-axis Attention (MCA). `MS' means multi-scale convolutions. `CA' means cross-axis attention. \revise{`$F$' denotes the original input feature.} We use MSCAN-T as the default encoder for this experiment.}
    \label{Ablation_mca}
    \begin{tabular}{lcccccccc} \toprule
    MS    &CA   &\revise{$F$}   & Params (M)   & Flops (G)  & mIoU & F1-score\\
    \midrule
    \xmark   & \xmark  & \revise{\xmark} &4.03   &6.94   &86.05  &89.78   \\
    \cmark   & \xmark  & \revise{\cmark}  &4.03   &7.29   &87.41  &90.76   \\
    \xmark   & \cmark  & \revise{\cmark}  &4.03   &7.23   &88.42  &91.39   \\
    \cmark   & \cmark  & \revise{\xmark}  &4.03   &7.29   &87.02  &90.27	  \\
    \midrule
    \cmark & \cmark & \revise{\cmark} &4.04  &7.30  &90.40  &92.93	  \\ 
    \bottomrule
    \end{tabular}
    \vspace{-2em}
\end{table}

\myPara{Importance of Each Component:}
We conduct experiments to evaluate the impact of multi-scale cross-axis attention on \nameofmethod{}.
The results are shown in Table~\ref{Ablation_mca}.
We can observe that when only the multi-scale convolutions is used, we receive improvement of 0.98 and 1.36 for the F1-score and mIoU, respectively.
When only cross-axis attention is used, we have improvement of 1.61 and 2.37 on the F1-score and mIoU, respectively.
When both of them are used, the F1 and mIoU can improve by 3.15 and 4.35, respectively.
These indicate that our multi-scale convolutions and cross-axis attentions are complementary.
Both multi-scale features and global context play essential roles in medical image segmentation.
\revise{The features $F$ act as a residual connection, which is similar to the one in the self-attention mechanism.
The goal is to accelerate training and improve training stability.
As can be seen from the table, the use of $F$ in MCA yields better performance.}

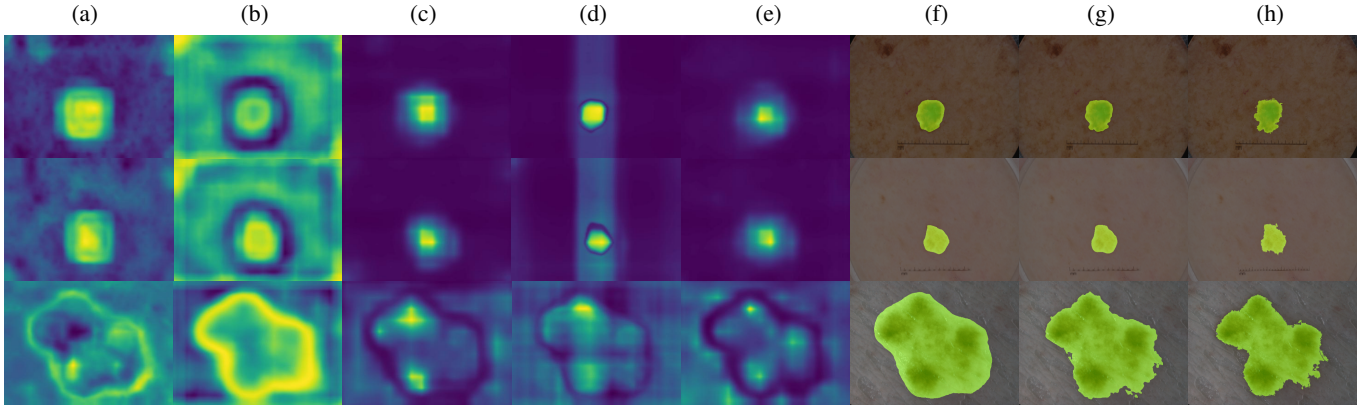
\begin{figure*}[tp]
  \small
  \centering
  \begin{overpic}[width=18cm]{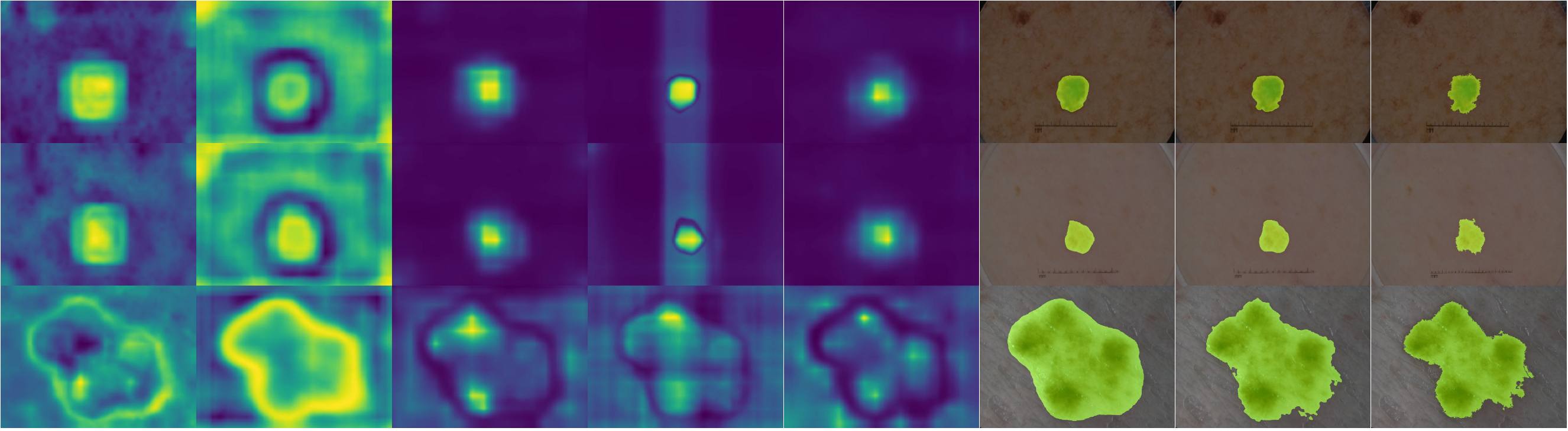}
    \put(5   ,28.5){{\textcolor{black}{(a)}}}
    \put(17.5,28.5){{\textcolor{black}{(b)}}}
    \put(30.0,28.5){{\textcolor{black}{(c)}}}
    \put(42.5,28.5){{\textcolor{black}{(d)}}}
    \put(55.5,28.5){{\textcolor{black}{(e)}}}
    \put(68.0,28.5){{\textcolor{black}{(f)}}}
    \put(80.0,28.5){{\textcolor{black}{(g)}}}
    \put(92.5,28.5){{\textcolor{black}{(h)}}}
  \end{overpic}
  \caption{Visual analysis of the feature maps produced by different versions of MCA. All sample are selected from the ISIC-2018 dataset. (a) w/o multi-scale cross-axis attention; (b) w/ multi-scale convolutions only; (c) w/ cross-axis attentions only;
  (d) axial attention; (e) w/ multi-scale cross-axis attention; (f) segmentation results w/ axial attention; (g) segmentation results w/ multi-scale cross-axis attention; (h) ground-truth annotations. \revise{In the visual analysis, we use the features before the output of~\nameofmethod{}.}}
  \label{fig:hotmap}
   \vspace{-1.5em}
\end{figure*}

To further investigate how each component works in MCA, we visualize the feature maps produced by different versions of MCA using the method proposed in~\cite{Zagoruyko2017AT}.
The results are shown in Fig.~\ref{fig:hotmap}.
Multi-scale convolutions are good at locating the rough lesion regions.
At the same time, cross-axis attention can help further identify the boundaries of the lesion regions from a global view.
Both of them are important to generate high quality segmentation results for medical images.

\begin{figure}[tp]
  \small
  \centering
  \begin{overpic}[width=0.49\textwidth]{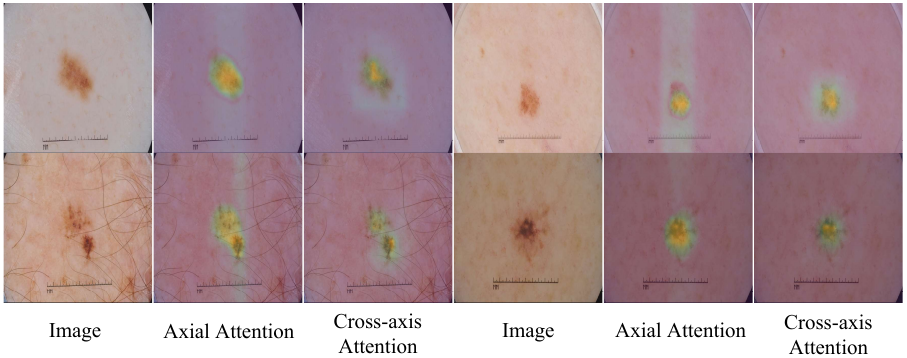}
  \end{overpic}
  \caption{\revise{Visual analysis of feature maps generated by ablation study on `$E_1$'. In the visual analysis, we use the features before the output of~\nameofmethod{}.}}
  \label{fig:hotmap-E1}
     \vspace{-1.3em}
\end{figure}

\myPara{Cross-axis Attention:}
We conduct experiments to evaluate the impact of the cross-axis attention.
The results of this experiment are shown in Table~\ref{Ablation_crossaxial}.
We use multi-scale convolutions by default.
We can see that using cross-axis attention shows a significant performance improvement with fewer parameters and computational complexity compared to using Axial Attention~\cite{ho2019axial}.
Compared to using two parallel axial attentions (w/o cross in CA), an improvement of 0.89 and 1.22 in F1 score and mIoU, respectively, is achieved.
%
Removing either cross-x-axis or cross-y-axis attention leads to significant performance drop.
%
As shown in Fig.~\ref{fig:hotmap}, multi-scale cross-axis attention performs much better than axial attention, especially when processing the lesion region boundaries.
\revise{In addition, we list the results of MedT~\cite{valanarasu2021medical} and sparse sinkhorn attention~\cite{tay2020sparse} that are similar to our method~\nameofmethod{}, and it can be observed from the table that \nameofmethod{} still outperforms them.}

\begin{table}[!tp]\centering
    \small
    \centering
    \setlength\tabcolsep{4pt}
    \caption{Ablation study on the cross-axis attention. `CA' means `cross-axis attention'. `w/o Cross in CA' means we do not use cross attention between the two branches of MCA. We use MSCAN-T as the default encoder for this experiment.}
    \label{Ablation_crossaxial}
    \begin{tabular}{lccccc} \toprule
    Decoder   & Params (M)  & Flops (G) & mIoU & F1-score\\
    \midrule
    w/o CA         &4.03   &7.29   &87.41  &90.76 \\
    Cross-$x$-Axis  &4.04 &7.23 &87.91 &91.61   \\ 
    Cross-$y$-Axis  &4.04 &7.23 &87.94 &91.68   \\ 
    w/o Cross in CA      &4.04 &7.30 &89.18  &92.04   \\ 
    Axial Attention~\cite{ho2019axial} &4.07 &7.47 &85.41  &89.27 \\
    \revise{sparse sink. att.~\cite{tay2020sparse}} &4.05 &7.20 &85.95 &89.68  \\
    \revise{MedT~\cite{valanarasu2021medical}}  &1.37  &1.95  &87.59  &89.48    \\
    \midrule
    CA   &4.04  &7.30   &90.40   &92.93	  \\ 
    \bottomrule
    \end{tabular}
\end{table}

\begin{table}[ht]
    \centering
    \setlength\tabcolsep{10pt}
    \caption{ \revise{Ablation study on `$E_1$' in MCA. `MCA w/ $E_1$' means to use  $E_1$, $E_2$, $E_3$, and $E_4$ to participate in the calculation of MCA, and then use the output of MCA to predict the final segmentation result.}}
    \begin{tabular}{lcccccccc} \toprule
        {} &Params   &Flops   &\multicolumn{2}{c}{ISIC-2018} \\  \cmidrule(lr){4-5}
        Methods & (M)   & (G)  & mIoU &  F1-score \\ \midrule
        MCANet w/o  $E_1$  &4.03 &5.09   &86.92  &90.37    \\
        MCA w/  $E_1$   &4.04 &5.66  &89.70  &92.45    \\
        MCANet   &4.04 &7.30   &90.40  &92.93    \\ \midrule
    \end{tabular}
    \vspace{-1.5em}
    \label{E1}
\end{table}

\myPara{\revise{The role of $E_1$ in MCA:}} \revise{We conduct experiments to evaluate the impact of $E_1$ on~\nameofmethod{}, and the results are shown in Table~\ref{E1}.}
\revise{For a more intuitive understanding, we also conduct visual analysis as shown in Fig~\ref{fig:hotmap-E1}.
From the table, it can be observed that `MCANet' and `MCA w/ $E_1$' can provide significant improvement. 
Combined with the visual analysis, this improvement is mainly attributed to including essential boundary cues in $E_1$.
We further compare the results of `MCANet' and `MCA w/ $E_1$' to explore whether all feature maps from the encoder (i.e. $E_1$, $E_2$, $E_3$ and $E_4$) should be directly sent to MCA.
It can be observed from the table that `MCANet' improves the mIoU and F1-score by 0.70 and 0.48, respectively, compared to `MCA w/ $E_1$'.
We argue that this is because the low-level features of $E_1$ are not suitable for capturing high-level semantics but more appropriate for use as local features to refine the boundary details of the lesion regions.
The visual analysis shows that the above performance improvement is mainly attributed to the fact that `MCANet' retains more edge details, which indicates that the use of $E_1$ in our MCANet is suitable.}

\myPara{Encoder and Decoder:}
We also conduct experiments to evaluate the impact of different encoders and decoders on \nameofmethod{}.
First, we test the impact of different decoders on \nameofmethod{}.
As shown in the top part of Table~\ref{Ablation_endocer}, we use four different decoders, including CCNet~\cite{huang2019ccnet}, EMA~\cite{li2019expectation}, Non-Local~\cite{wang2018non}, and Hamburger~\cite{wang2018non}, for this experiment.
All of them are lightweight.
The results show that our proposed method performs better than other decoders in terms of both F1-score and mIoU.  
Furthermore, we test the effect of different encoders on~\nameofmethod{}.
We report results on commonly used encoders, such as ResNet-50~\cite{he2016deep}, MiT-B1~\cite{xie2021segformer} and Swin-T~\cite{liu2021swin}.
As shown in the bottom part of Table~\ref{Ablation_endocer}, it can be observed that using other popular encoders also results in good segmentation performance, but MSCAN-T performs the best.
%
To conclude, we can see that our proposed \nameofmethod{} achieves the best trade-off among the number of parameters, computational complexity, and performance.



\begin{table}[!tp]\centering
  \small
  \centering
  \setlength\tabcolsep{1pt}
  \caption{Ablation study on encoder and decoder.}\label{Ablation_endocer}
  \begin{tabular}{lccccc} \toprule
Encoder  & Decoder  & Params (M)   & Flops (G)   & mIoU   & F1-score\\
\midrule
MSCAN-T  &Non-Local~\cite{wang2018non} &11.51   &6.67     &88.05  &91.83 \\
MSCAN-T  &CCNet~\cite{huang2019ccnet}  &11.31   &6.70     &88.33  &91.91 \\   
MSCAN-T  &EMA~\cite{li2019expectation} &7.71    &5.76     &89.29  &92.18 \\ 
MSCAN-T  &Hamburger~\cite{geng2021attention} &4.20  &6.60  &89.72  &92.12 \\ 
\midrule
ResNet-50    & MCA (Ours)    &24.41 &36.70   &86.33	&89.94	 \\ 
Swin-T       & MCA (Ours)    &27.67 &26.93   &88.08 &91.27   \\ 
MiT-B1       & MCA (Ours)    &13.71 &16.71   &88.52	&91.59	 \\ 
\midrule
MSCAN-T & MCA (Ours)      &4.04   &7.30       &90.40   &92.93	 \\ 
\bottomrule
     \vspace{-2em}
\end{tabular}
\end{table}

\section{Conclusion}
We propose \nameofmethod{} for medical image segmentation.
The core component of \nameofmethod{} is the multi-scale cross-axis attention, a new method that combines multi-scale features and cross-axis attention to better segment organs or lesions with different sizes and shapes. 
Extensive experiments on four typical medical image segmentation tasks show that the proposed \nameofmethod{} outperforms previous state-of-the-art methods with fewer parameters and lower computational complexity.
%


\bibliographystyle{IEEEtran}
\bibliography{egbigb}

\end{document}